\documentclass[reprint,prx,notitlepage,superscriptaddress]{revtex4-1}

\usepackage{graphics,epsfig,amsfonts,amssymb,amsmath,color}
\usepackage[normalem]{ulem}
\usepackage{bm,bbm}
\usepackage[mathcal]{euscript}
\usepackage[monochrome]{xcolor}
\usepackage{soul}
\usepackage{float}

\newcommand{\ket}[1]{\ensuremath{|{#1}\rangle}}
\newcommand{\braket}[2]{ \langle #1 | #2 \rangle }

\colorlet{BLUE}{blue}

\begin{document}

\title{Mirror symmetry breaking of superradiance in a dipolar BEC}

\author{Bojeong Seo}
\thanks{These authors contributed equally.}
\affiliation{Department of Physics, The Hong Kong University of Science and Technology, Clear Water Bay, Kowloon, Hong Kong, China}

\author{Mingchen Huang}
\thanks{These authors contributed equally.}
\affiliation{Department of Physics, The Hong Kong University of Science and Technology, Clear Water Bay, Kowloon, Hong Kong, China}

\author{Ziting Chen}
\affiliation{Department of Physics, The Hong Kong University of Science and Technology, Clear Water Bay, Kowloon, Hong Kong, China}

\author{Mithilesh K. Parit}
\affiliation{Department of Physics, The Hong Kong University of Science and Technology, Clear Water Bay, Kowloon, Hong Kong, China}

\author{Yifei He}
\affiliation{Department of Physics, The Hong Kong University of Science and Technology, Clear Water Bay, Kowloon, Hong Kong, China}

\author{Peng Chen}
\affiliation{Department of Physics, The Hong Kong University of Science and Technology, Clear Water Bay, Kowloon, Hong Kong, China}

\author{Gyu-Boong Jo}
\email{gbjo@ust.hk}
\affiliation{Department of Physics, The Hong Kong University of Science and Technology, Clear Water Bay, Kowloon, Hong Kong, China}
\affiliation{IAS Center for Quantum Technologies, The Hong Kong University of Science and Technology, Kowloon, Hong Kong, China}

\begin{abstract}
Dicke superradiance occurs when two or more emitters cooperatively interact via the electromagnetic field. This collective light scattering process has been extensively studied across various platforms, from atoms to quantum dots and organic molecules.
Despite extensive research, the precise role of direct interactions between emitters in superradiance remains elusive, particularly in many-body systems where the complexity of interactions poses significant challenges.
In this study, we investigate the effect of dipole-dipole interaction between 18,000 atoms in dipolar Bose-Einstein condensates (BECs) on the superradiance process. In dipolar BECs, we simplify the complex effect of anisotropic magnetic dipole-dipole interaction with Bogoliubov transformation. We observe that anisotropic Bogoliubov excitation breaks the mirror symmetry in decay modes of superradiance.
 \end{abstract}

\maketitle

Collective light scattering~\cite{Dicke.1954je7}, a cooperative emission process inducing directional scattered atoms, has been observed in various atomic systems ranging from thermal atoms~\cite{Gross.1982l7,Yoshikawa.2005},  degenerate Bose gases~\cite{Inouye.1999,Stenger.1999rlo,Inouye.19990n6,Fallani.2005,Deng.2010gfl,Lu.2011je5,Kampel.2011,Lopes.2014,Dimitrova.2017}, free fermions~\cite{Wang.2010f8xf} to atoms coupled to the cavity mode~\cite{Slama.2007,Baumann.2010,Kessler.2014}.
Among them, a Bose-Einstein condensate (BEC) has served as a promising platform for exploring a superradiant light scattering process owing to its unique coherence property with~\cite{Inouye.1999,Stenger.1999rlo,Inouye.19990n6,Fallani.2005,Deng.2010gfl,Lu.2011je5,Kampel.2011,Lopes.2014,Dimitrova.2017,Deng.2010} and without external light fields~\cite{Clark.2017,Fu.2018i3f,Wu.2019p5j,Kim.2021}.
When the external light shines on atoms in the condensate, collective scattering of light creates a quasiparticle in the form of recoiling atoms that interfere with condensate atoms at rest, leading to the generation of matter-wave grating that is further enhanced by subsequent light scattering~\cite{Muller.2016}. So far, however, prior studies have primarily focused on a non-interacting regime, leaving the effect of interactions largely unexplored ~\cite{Gross.1982,Lahaye.2009,Bismut.2012,Wenzel.2018}. This makes it challenging to explore exotic states of matter using light scattering even though fundamental excitations in exotic states are expected to manifest themselves in the light-matter interaction process~\cite{Deng.2013}.

\vspace{10pt}
In this work, we investigate a superradiance light scattering process from a dipolar BEC in an elongated trap, in which the mirror symmetry in superradiance is broken due to the dipolar intearction. Here, magnetic dipolar interactions result in anisotropic dispersion of dipolar superfluid~\cite{Bismut.2012,Wenzel.2018}, which allows us to control the asymmetry of superradiant peaks by changing the direction of the external magnetic field. Such dipolar effects in quantum gases~\cite{Chomaz.2022} have recently opened up a new regime where anisotropic dipole-dipole interactions play a crucial role in realizing new phases of matter, such as quantum droplets and supersolids~\cite{Chomaz.2022}. Our work demonstrates how fundamental excitations begin to contribute and manifest themselves in light-matter interaction processes.

\begin{figure}
	\includegraphics[width=\linewidth]{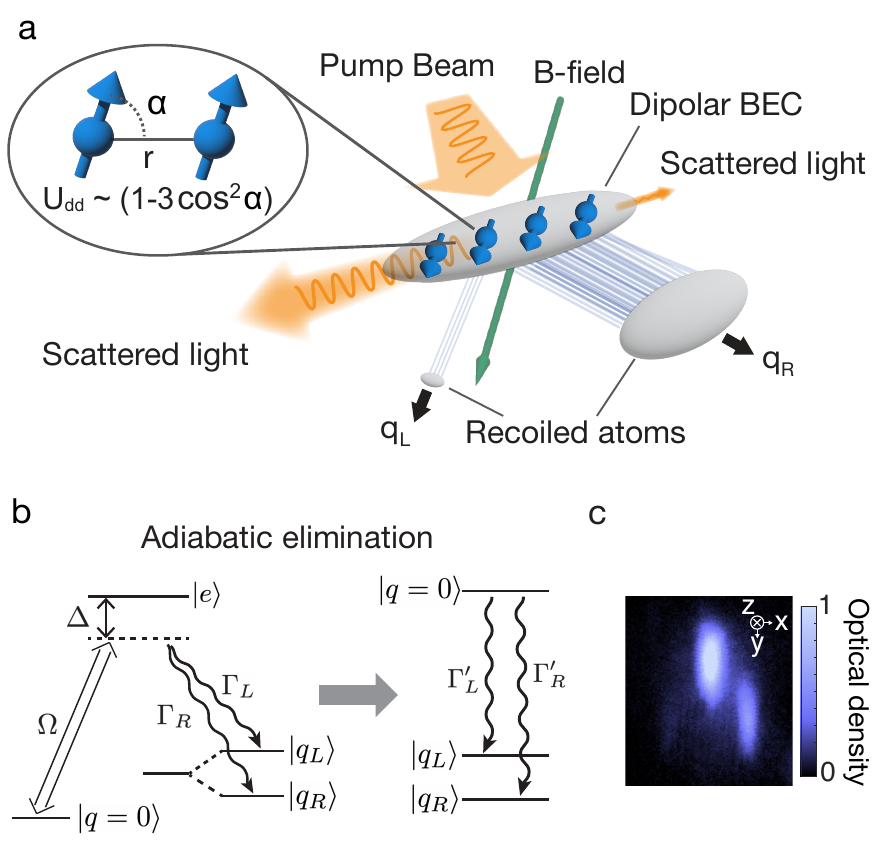}
	\caption{{\bf Mirror symmetry breaking of the superradiance in a dipolar BEC} (a) A quasi-1D dipolar condensate is exposed to the pump beam propagating along the $+$y axis. Due to the anisotropic dipolar interaction presented in the dipolar BEC (circular inset), the Bogoliubov excitation spectrum becomes anisotropic, leading to the mirror symmetry breaking for two exciations along the elongated direction of the condensate. As a result, we observe the asymmetric outcoupled atoms after the expansion of the cloud. (b) The superradiance process can be simplified by adiabatically eliminating the intermediate state $\ket{e}$ for a sufficiently large detuning $\delta=6.6~MHz$, compared to the optical linewidth. Given the anisotropic dispersion spectrum, the two final ground states, $\ket{q_R}$ or $\ket{q_L}$, are shifted in energy with different transition rates. (c)  Time-of-flight image of asymmetric superradiant dipolar BEC.}
	\label{fig1}
\end{figure}

\vspace{10pt}
\paragraph*{\bf Experiment} We initiate experiments with a quasi-one-dimensional dipolar BEC of approximately $1.8 \times 10^4$ erbium ($^{168}$Er) atoms in the $\ket{M_J=-6}$ state~\cite{Seo.2020,qrl,He.2024}. By aligning the dipole orientation along a specified direction (parallel to an external magnetic field), we excited the ground state BEC at \textcolor{blue}{$\ket{q=0}$} to an excited state $\ket{e}$ using a largely detuned pump beam. This excited state then decays back to the ground state, acquiring non-zero momentum in the process due to momentum conservation. In a typical condensate without dipolar interactions, this scattering process is amplified along the long axis of the condensate~\cite{Inouye.1999, Stenger.1999rlo, Inouye.19990n6}. It leads to decay in two opposite directions symmetrically and emits photons with momentum $\hbar{k}_{583}$, where $k_{583}$ represents the wavevector of the pump beam. As a result of this symmetric light-scattering process, the quasi-particles are created with the momentum of $\sqrt{2}\hbar{k}_{583}$ along the 45$^\circ$ relative to the long-axis of the condensate ((see Fig.~\ref{fig1}(a)). We can simplify our model by eliminating the intermediate excited state and considering a two-level system comprising the initial state at \textcolor{blue}{$\ket{q=0}$} and a final state at \textcolor{blue}{$\ket{q=\sqrt{2}\hbar k_{583}}$}, owing to the relatively large detuning ($\Delta$) of the pump beam compared to the Rabi frequency ($\Omega$) of the driving field and the spontaneous emission rate from $\ket{e}$ to \textcolor{blue}{$\ket{q=k_{583}}$}.

\vspace{10pt}
A distinct feature of a dipolar BEC, as opposed to one with only isotropic contact interactions, is its anisotropic dispersion relation due to dipole-dipole interactions. The elementary anisotropic dipolar Bogoliubov excitation spectrum, \textcolor{blue}{$\hbar\omega(q)$}, for a uniform density $n$ is expressed as
\textcolor{blue}{\begin{equation}
    \hbar\omega(q)=\sqrt{E(q)(E(q)+2gn(1+\epsilon_{dd}(3\cos^2{\phi}-1)))}
    \label{eq1}
\end{equation}}
where \textcolor{blue}{$E(q)=\frac{\hbar^2q^2}{2m}$} and $\phi$ is the angle between the external magnetic field and the atomic excitation direction. Here, $g=4\pi \hbar^2 a_s/m$ with atomic mass $m$ and $\epsilon_{dd}=a_{dd}/a_s$ for the characterisitc dipolar length of $a_{dd}$=66.3$a_0$. This leads to a broken symmetry in the two end-fire modes, as evident from the asymmetry in their decay rates, ${\Gamma_L'}\neq{\Gamma_R'}$ (see Fig.~\ref{fig1}(b)). In consequence, one decay channel becomes dominant in the scattering process, leading to an observed increase in atoms in one of the two collective excitation (see Fig.~\ref{fig1}(c)).


\begin{figure}[t] 
	\includegraphics[width=\linewidth]{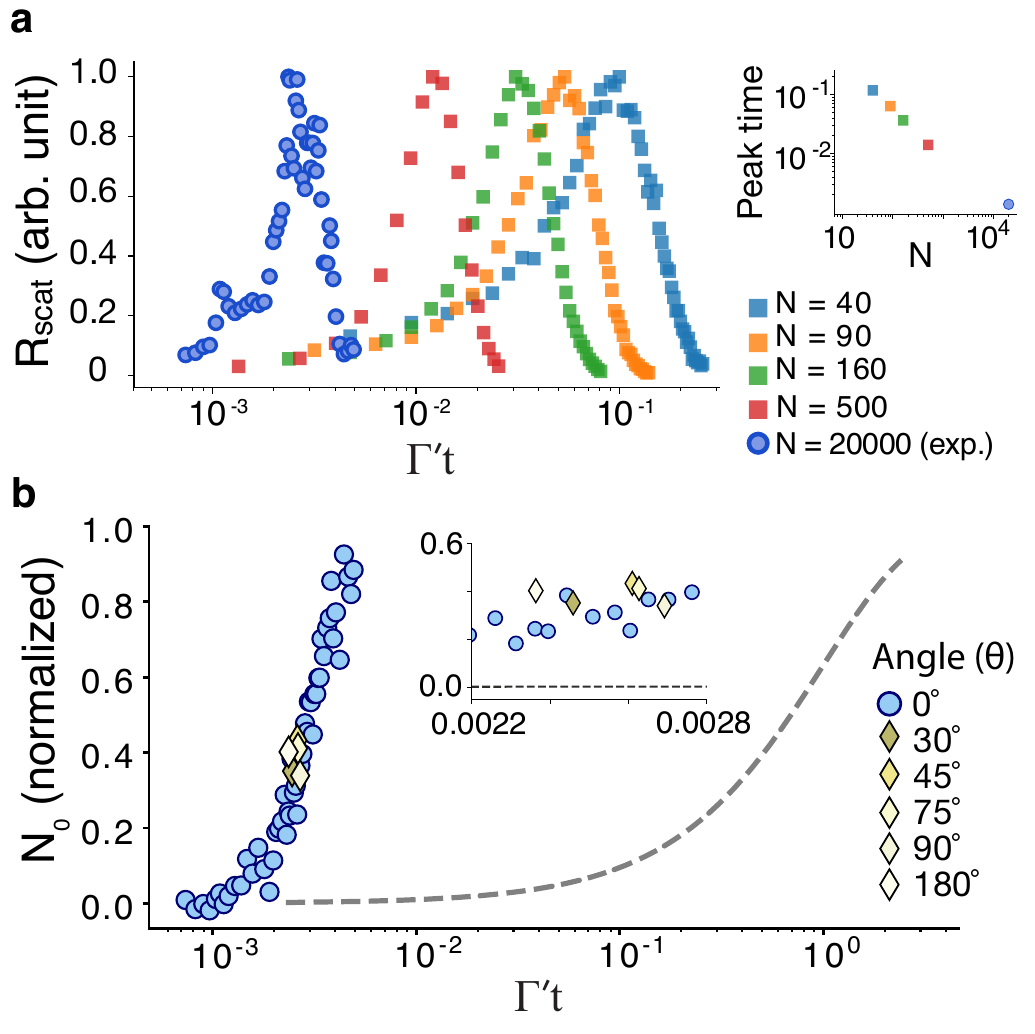}
	\caption{{\bf Collective enhancement of the atom-light scattering in dipolar BEC} (a) Bursts in the dicke superradiance with the measurement and the simulation. Data is normalized such that the peak $R_{scat}$ are set to 1. Inset shows the time of peak scattering for different number of atoms. (b) Change of the ground state population ($N_0$) against the change of $\Gamma't$. The dashed line shows the calculated single atom Raman scattering. $N_0$ is normalized such that maximum $N_0$ is 1. }
    \label{fig2}
\end{figure}


\begin{figure*}[t]
	\includegraphics[width=1.\textwidth]{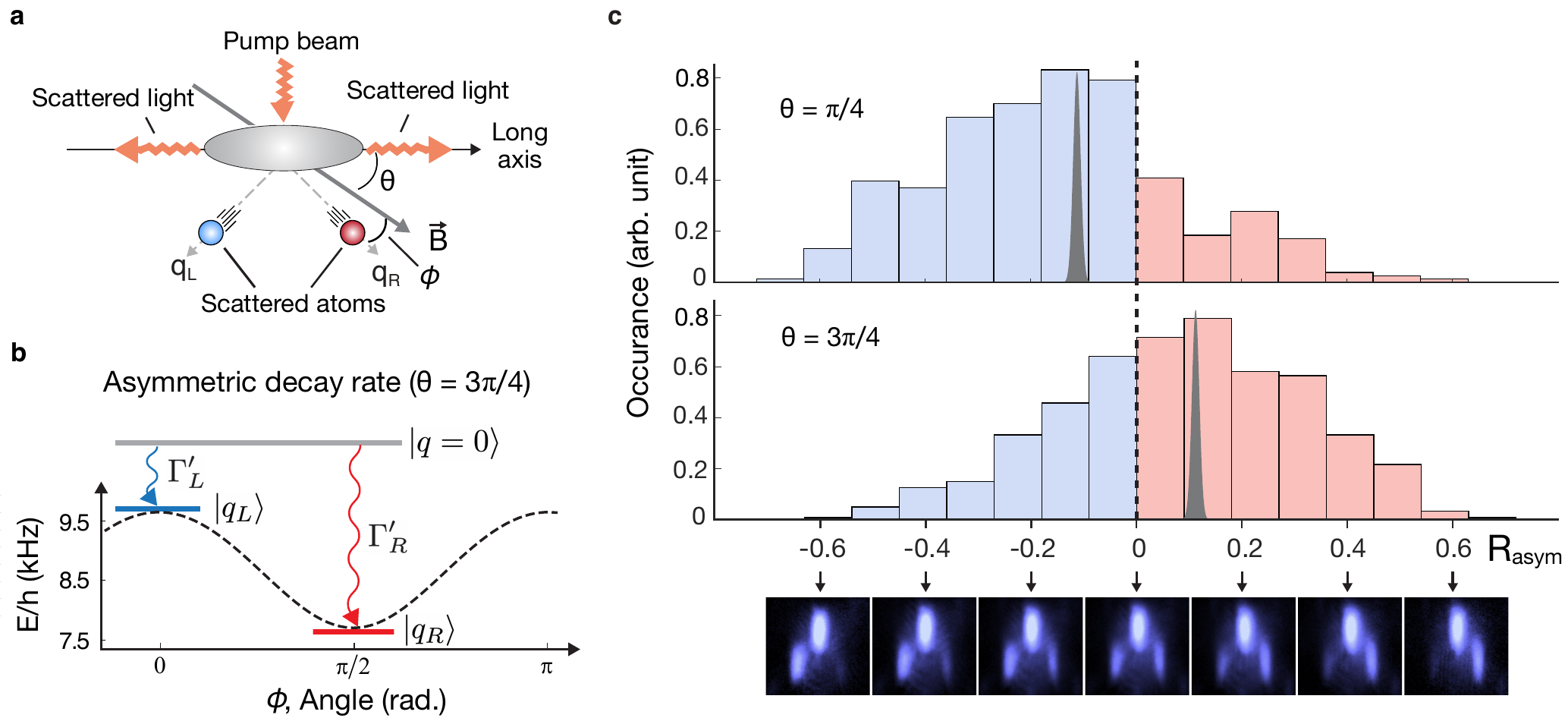}
	\caption{ {\bf Experimental demonstration of asymmetric superradiance in a dipolar BEC.} (a) Geometry of asymmetric excitation spectrum in experimental setup. Orange arrows represent the pump beam (top) and scattered lights (side). Grey ellipse represents the original dipolar condensate. Blue (red) circle represents the atoms transferred to \textcolor{blue}{$\ket{q_L}$ ($\ket{q_R}$)} state. (b) This is exemplified by an asymmetric dipolar Bogoliubov excitation spectrum in Eq.~(\ref{eq1}). At $\theta=\pi/4$, where the decay rates for leftward ($\Gamma'_L$) and rightward ($\Gamma'R$) excitations vary because the leftward excitation's energy level, $\ket{q_L}$, is higher than that of the rightward excitation, $\ket{q_R}$.  (c) By counting the number of atoms in $\ket{q_L}$ ($\ket{q_R}$) states for each experimental sequence, the distributions of $R_{asym}$ are obtained for $\theta=\pi/4$ and $3\pi/4$. Blue (red) regime correspond to $q_L$ ($q_R$) dominant data. \textcolor{red}{Grey distributions theoretically result from assuming independent light scatterings. } (low) Averaged adsorption images after 16ms TOF for corresponding $R_{asym}$ vlalues are provided.}
	\label{fig3}
\end{figure*}

\vspace{10pt}
To confirm that the observed asymmetric scattering is a collective phenomenon and not typical spontaneous Raman scattering, we measure the dynamic change in the scattering rate of atom decay. In Fig.\ref{fig2}(a), we observe the burst in scattering, which is consistent with quantum Monte-Carlo simulations, confirming the distinct collective nature of this interaction. To simulate this behavior, we solve the Lindblad master equation,
\begin{equation}
    \rho(t) = -\frac{i}{\hbar}[H, \rho(t)] + \sum_{i} \frac{\Gamma_i}{2} L(\rho),
\end{equation}
that describes the superradiant decay rate of \(N\) atoms with multiple decay modes, \(\Gamma_i\). The decay rate can be estimated by taking the time derivative of the population of the ground state. Assuming two decay modes, \(\Gamma_L\) and \(\Gamma_R\), the Lindblad equation can be simplified to $\rho(t) = \frac{\Gamma_L}{2} L(\rho) + \frac{\Gamma_R}{2} L(\rho)$. We use a quantum Monte Carlo solver in the Quantum Toolbox in Python (QuTiP)~\cite{qutip} to solve this equation.

As more atoms participate in the collective emission process, the peak scattering time advances, as shown in Fig.\ref{fig2}.a and its inset. In Fig.\ref{fig2}(b), we also confirm enhanced decay to the ground state  in asymmetric cases ($\theta=\pi/6, \pi/4$, and $5\pi/12$), indicating superradiance even when mirror-symmetry is broken. The ground state population resulted from the typical spontaneous Raman scattering without collective scattering is provided as a comparison (dashed line).

\vspace{10pt}
{\bf Mirror-symmetry breaking in superradiance} Next we examine the mirror-symmetry breaking mechanism with the statistical analysis of our measurements.
In Fig.~\ref{fig3}, we characterize a superradiant sample by recording an asymmetry of superradiant peaks. For each condition, we obtain around 170-300 data and plotted a histogram for an asymmetry ratio ($R_{asym}=\frac{N_R-N_L}{N_R+N_L}$) where $N_R$ ($N_L$) represents the atom number of right (left) superradiant peak. When the left (right) atom cloud is dominant, the mean value of the histogram has a negative (positive) value.

\vspace{10pt}
Scattering events create two quasi-particle excitations, $k_L$ and $k_R$, with different angles ($\phi$) relative to the external magnetic field, leading to asymmetric decay rates (see Fig.~\ref{fig3}(a) and (b)). For instance, at $\theta=3\pi/4$, the decay rate to $\ket{q_R}$ ($\Gamma_R'$) is stronger than the decay rate to $\ket{q_L}$ ($\Gamma_L'$) since the detuning for the transition from $\ket{q=0}$ to $\ket{q_R}$ is smaller than $\ket{q_L}$. As a result, the population in the $q_R$ mode is predominantly observed (see Fig.~\ref{fig3}(c)). Histograms of asymmetry ratio, $R_{asym}$, from the data in Fig.~\ref{fig3}(c) show this dominance. It is important to note that a non-negligible portion of the data exhibits dominant excitation to $q_L$ ($q_R$) even when $\Gamma_R' > \Gamma_L'$ ($\Gamma_L' > \Gamma_R'$) at $\theta=3\pi/4$ ($\theta=\pi/4$). This result is due to the stochastic nature of the initial scattering events, which trigger an avalanche-like scattering of the remaining atoms. This outcome aligns with the collective behavior observed in Fig.2. 
Similar broken-mirror symmetry in superradiance has recently been studied theoretically in chiral-waveguide system~\cite{Cardenas-Lopez.2023}, where a similar wide distribution of $R_{asym}$ is predicted. We note that the theoretical estimation for the chiral-waveguide system expected that peak of the histogram is closer to $R_{asym}=-1$ or $1$. We attribute the difference in our experimental data to several experimental imperfections, such as decoherence or non-uniform distribution of atoms in the dipolar BEC.

\begin{figure}
	\includegraphics[width=0.43\textwidth]{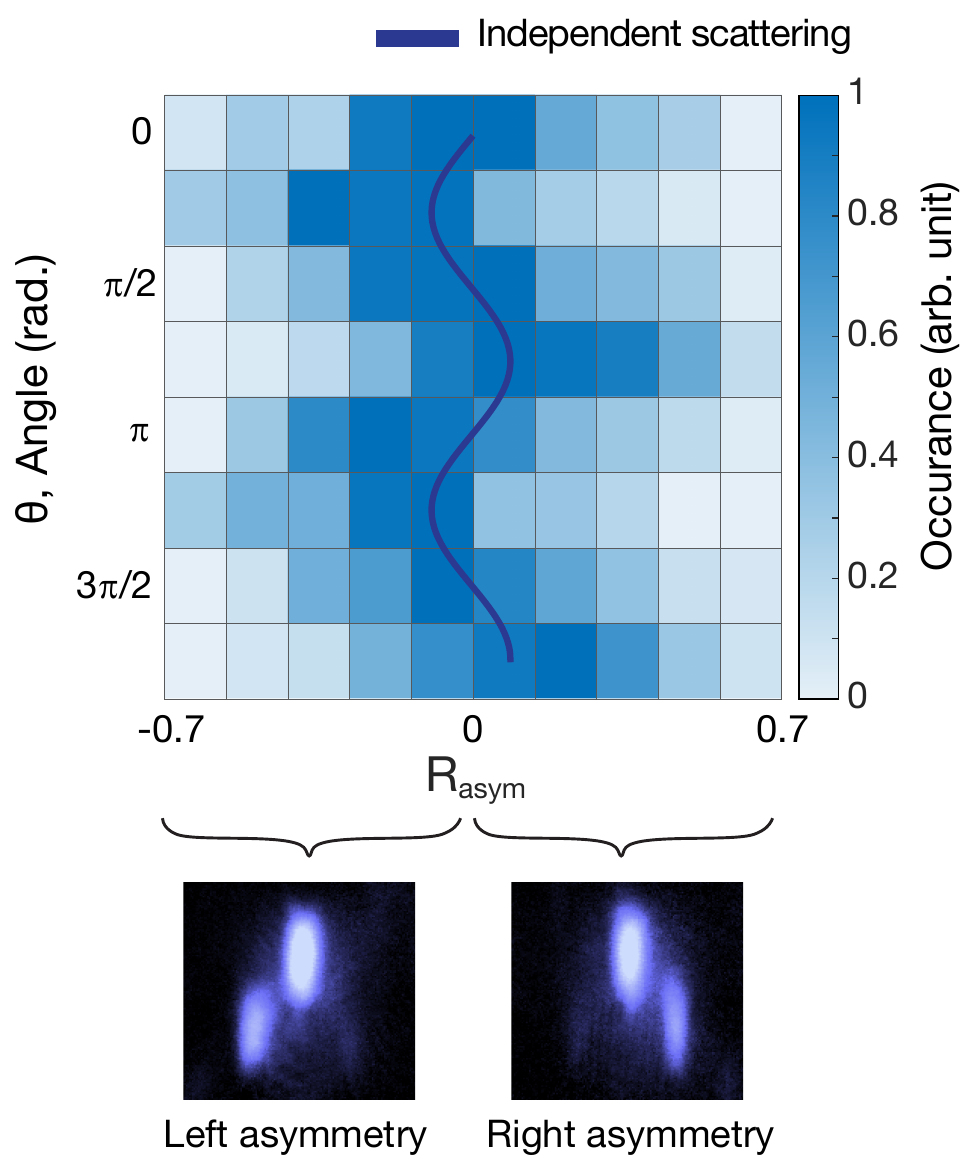}
	\caption{{\bf Observation of asymmetric superradiant peaks}. \textcolor{red}{Heatmap of distribution of $R_{asym}$ against the change of the orientation of the magnetic dipoles in the dipolar BEC. Solid blue line is obtained from the independent light scattering model for ${\Gamma_L'}\neq{\Gamma_R'}$. Distribution of each angle is normalized such that the peak occurrence is set to 1. Momentum distribution of BECs after 16~ms time-of-flight are shown below. left (right) asymmetric recoiling atoms corresponding to the negative (positive) mean value of $R_{asym}$}}
	\label{fig4}
\end{figure}

\vspace{5pt}
{\bf Angle dependence of asymmetric superradiance} Lastly, we further examine the angle dependence of symmetry breaking in  superradiant scattering by varying the angle of the dipole orientation, $\theta$, from $0$ to $7\pi/4$. Fig.~\ref{fig4}. depicts the measured periodic oscillation of asymmetry against the change of dipole orientation. To elucidate the effect of magnetic field direction on superradiance, we keep the Rayleigh scattering rate constant and therefore effectively the gain term of superradiance the same. In our experiment, the end-fire mode always occurs along the x direction regardless of the magnetic field direction. In this case, the spontaneous emission strength along the x direction is sensitive to the quantization direction (i.e. external magnetic field) following the radiation pattern (see supplementary note). Therefore, we calibrate the intensity of the pump beam such that spontaneous emission along the longitudinal direction of the atom cloud remains constant for the different quantization directions. Furthermore, to avoid the shape change of condensate induced by the field direction we switch on a pump beam immediately after rapidly rotating the magnetic field in less than 0.5~ms.

 When the magnetic field is oriented at $\theta$=0$^{\circ}$ or 90$^{\circ}$, atomic excitations that involve opposite matter-wave gratings occur with the same strength as $\theta_k$ is $\pi/4$ for both cases resulting in $\Gamma_R=\Gamma_L$. The effect of anisotropic dispersion becomes prominent when the magnetic field is aligned to one of the excitation directions (e.g. $\theta$=45$^{\circ}$ or 135$^{\circ}$). At $\theta$=45$^{\circ}$, for example, the atomic excitation involving the right superradiant peak ($N_R$) undergoes stiffer dispersion with a larger excitation resonance frequency (for $\theta_q$=0) compared to the left peak ($N_L$). According to the f-sum rule~\cite{Nozieres.1990}, the excitation involving the increased resonance frequency (i.e. right peak) is relatively more suppressed than the left peak with a smaller resonance frequency with $\Gamma_R\neq\Gamma_L$ (see Fig.~\ref{fig4}).

\vspace{10pt}
Our work opens an interesting avenue for future experiments. First, when a dipolar condensate is exposed to the pump beam across the BEC to the quantum droplet~\cite{petrov.2015,Ferrier.2016,chomaz.2016}, it would be worthwhile to explore how unusual states of matter appear in the process of interaction between light and matter~\cite{Huang.2023}. The coherence property of quantum droplet can be further manifested by the superradiance light scattering. Another intriguing idea is to study how the emission direction can be spontaneously broken with shot-to-shot fluctuation associated with avalanche-like behavior of superradiance~\cite{Cardenas-Lopez.2023}. Full characterization of fluctuation may allow us to explore memory effects in the presence of dipolar and/or s-wave interactions~\cite{Cardenas-Lopez.2023}.

In conclusion, we have explored a superradiant process in a dipolar BEC and explained how magnetic dipolar interactions influence atom-light scattering. The anisotropic dispersion allows us to manage asymmetric superradiant peaks by adjusting the external magnetic field's direction. Future research could explore a superradiant process in an oblate dipolar condensate~\cite{Ticknor.2011}. This could potentially control the direction of recoiling atoms based on the external magnetic field direction. Another interesting area to explore is collective light scattering in dipolar fermions~\cite{Lu.2012,Aikawa.2014} or dipolar molecules~\cite{Ni.2008}.

\vspace{10pt}
\paragraph*{\bf Acknowledgements} We thank Ana Asenjo Garcia, Stuart Masson, Eric Sierra, and Hui Zhai for fruitful discussions. We acknowledge support from the RGC through  16306119, 16302420, 16302821, 16306321, 16306922, 16302123, C6005-17G, C6009-20G, and RFS2122-6S04. This work is also supported by the Guangdong-Hong Kong Joint Laboratory and the Hari Harilela foundation.

\renewcommand{\thefigure}{S\arabic{figure}}
\setcounter{figure}{0} 

{\large \bf Supplementary Note}
\vspace{20pt}

\paragraph*{\bf Experimental details}
Our experiments begin with $^{168}$Er BEC of approximately 1.8(2)$\times 10^4$ atoms in the $\ket{M_J=-6}$ state~\cite{Seo.2020,qrl}. The atoms are trapped in the crossed optical dipole trap (ODT) consisting of two 1064~nm laser beams propagating along the x-direction and y-direction with the beam waist of $w_{y-z}$=20~$\mu m$ and $w_{x-z}$=45~$\mu m$, respectively. To observe superradiance, we gradually change the trap geometry over 45~ms, resulting in a quasi-1D trap with the trap frequency of $(\omega_x,\omega_y,\omega_z)=2\pi\times(37, 445, 443)~$Hz and the maximum chemical potential of $\mu\simeq h\times$ 3.5~kHz at $T\simeq$~140(10)~nK. During evaporative cooling, the magnetic field remains at 400~mG along the y-direction. Before pulsing a pump light, we adjust the magnetic field to the target value at a variable angle over $t_{ramp}$ and switch off all ODT's within 60~$\mu$s. At this stage, phase fluctuation is controlled by the magnitude of bias magnetic field $B_0$ (near the Feshbach resonance) while the anisotropic dispersion relation is set by the field direction $\theta$ as described in the main text.
 
 To induce superradiant Rayleigh Scattering, an elongated condensate is illuminated with a single off-resonant 583~nm laser beam for 100~$~\mu$s. The pump beam,  linearly polarized along the z-direction, is about 100 times larger than the size of the sample - with the Thomas-Fermi radius of 15~$\mu$m - to maintain the same optical intensity along the condensate. Additionally, the incident angle of the pump beam is perpendicular to the elongated axis to circumvent  the unintended asymmetric superradiance~\cite{Lu.2011je5}. Atomic momentum distribution after collective light scattering is recorded after a 16~ms of time-of-flight expansion using absorption imaging with a circularly polarized ($\sigma^-$) 401~nm  broad-linewidth transition ($4f^126s^2 ~{ }^3H_6~\rightarrow~4f^{12}({}^3H_6)6s6p({}^1P_1)$) along the z-direction.  With data from the absorption imaging, we quantify the number of recoiled atoms. The total number of recoiled atoms is denoted as ${N=N_L+N_R}$ where $N_L$ and $N_R$ are the number of atoms in left and right recoiled atom clouds, respectively.

\vspace{.05in}
\paragraph*{\bf Scattering rate of atoms}
Scattering rate, $\Gamma$, is calculated using Fermi's Golden Rule, which relates the transition rate to the square of the matrix element of the interaction Hamiltonian,
\begin{equation}
    \Gamma_{i{\rightarrow}f}=\frac{2\pi}{\hbar}|\braket{f|\hat{H'}}{i}|^2\delta(E_f-E_i).
    \label{eq:Fermi Golden}
\end{equation}
The perturbation Hamiltonian, $\Hat{H'}$, for the coupling between the light field and the atoms is interaction Hamiltonian,
\begin{equation}
    \hat{H}_{int}=-\int d^3r \, \hat{\bold{\Psi}}^\dagger(\bold{r}){\Hat{\bold{d}}}\cdot{\Hat{\bold{E}}(\bold{r})}\hat{\bold{\Psi}}(\bold{r})\,
    \label{eq:H_int1}
\end{equation}
where $\hat{\bold{\Psi}}$ is the field operator for the atom in BECs, $\hat{\bold{d}}$ is the dipole moment, $\hat{\bold{E}}(\bold{r})$ is the electric field of the light field at $\bold{r}$.
The field operator for atom in the BECs is
\begin{equation}
    \hat{\bold{\Psi}}(\bold{r})=\sum_\bold{q}{\phi_\bold{q}}(\bold{r})a_\bold{q}.
\end{equation}
Here $a_\bold{q}$ are the annihilation operators for atoms in the momentum states with $\bold{q}$ and $\phi_\bold{q}(\bold{r})$ are the wavefunctions for the momentum states with $\bold{q}$. In this study, all traps are switched off while the light scattering occurs and BECs are in the free space. In free space, BECs can be ideally described by macroscopic plane waves,
\begin{equation}
    \phi_{\bold{q}}(r)=\frac{1}{\sqrt{V}}e^{i\bold{q}\cdot\bold{r}}.
    \label{eq:planewave}
\end{equation}
In turn, the interaction Hamiltonian in Eq.~\ref{eq:H_int1} for BECs in free space in second quantization is,
\begin{equation}
    \hat{H}_{int}=\sum_{\bold{q_i,q_f}}{g(\bold{k}_0+\bold{q}_i-\bold{q}_f)\hat{a}_\bold{q_i}\hat{a}^\dagger_\bold{q_f}}\hat{c}^\dagger_{\bold{k_0}+\bold{q_i}-\bold{q_f}}+h.c.
    \label{eq:H_int2}
\end{equation}
where $g(\bold{k})$ is the coupling to the light field and ${\hat{c}^\dagger}_k$ is the photon creation operator. Note that Eq.~\ref{eq:H_int2} includes annihilation and creation operators for atom in $\bold{q_i}$ and $\bold{q_f}$ originated from the plane wave approximation in Eq.~\ref{eq:planewave}. We need to consider both initial momentum state and the final momentum state, due to the momentum transfer from photon to atom.

\vspace{.05in}
\paragraph*{\bf Bogoliubov description for dipolar BECs}
Dipolar BECs have s-wave contact interaction as well as the long-range dipolar interaction. It is important to properly address these complex interactions between particles when study the coupling between light field and the BECs. To understand the low excitation of the dipolar BECs, we introduce the Bogoliubov theory. In Bogoliubov theory, Hamiltonian in second quantization can be diagonalized using the Bogoliubov transformation,
\begin{equation}
    \begin{aligned}
    \hat{a}_q=u_q\hat{b}_q+v_{-q}\hat{b}^\dagger_{-q}+\delta(q)\\
    \hat{a}^\dagger_q=u_q\hat{b}^\dagger_q+v_{-q}\hat{b}_{-q}+\delta(q).
    \end{aligned}
    \label{eq:Bog_transform}
\end{equation}

\begin{figure}
	\includegraphics[width=\linewidth]{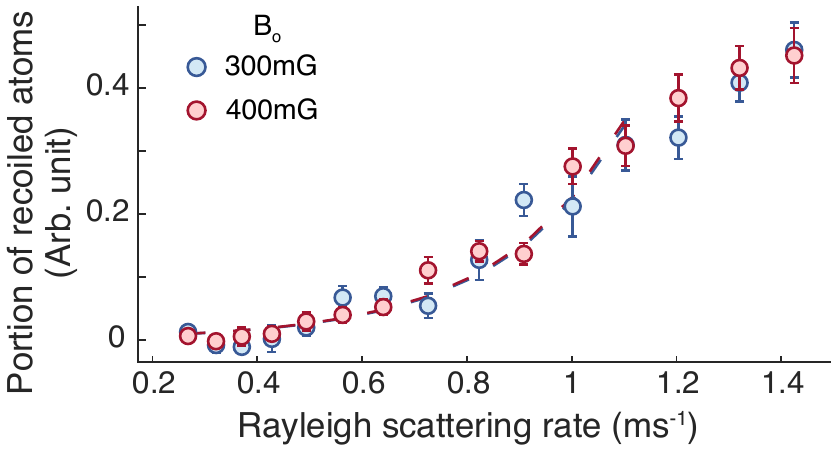}
	\caption{ {\bf Rayleigh scattering rate calibration.} The Rayleigh scattering rate is independently controlled at different magnetic fields.  For example, we calibrate the superradiance at  300~mG and 400~mG, where $a_s$ are almost the same. The dashed lines are a guideline. The error bar is the standard deviation.}
	\label{figS1}
\end{figure}

Here, $\hat{b}_q$ and $\hat{b}^\dagger_q$ are the Bogoliubov excitation annihilation and creation operators. Also, the Bogoliubov coefficients $u_q$ and $v_q$ are
\begin{equation}
    u_q^2, v_{-q}^2 = \frac{1}{2}\left[ \frac{\epsilon_q+nV_{eff}}{E(q)}\pm1\right]
\end{equation}
where n is the density of the BEC and $V_{eff}$, the effective interaction potential, for dipolar BEC is given by
\begin{equation}
    V_{eff}=g_{contact}(1+\epsilon_{dd}(3cos^2\theta_q-1)),
\end{equation} 
$E(q)$ is the Bogoliubov dispersion law for the dipolar Bogoliubov excitation specturm,
\begin{equation}
 E(q)=\left[\epsilon_q\left(\epsilon_q+2nV_{eff}\right)\right]^{1/2},
\end{equation}
and $\epsilon_q$ is the kinetic energy,
\begin{equation}
    \epsilon_q=\frac{\hbar^2q^2}{2m}.
\end{equation}
Since the Bogoliubov coefficients, $u_q$ and $v_q$ diverge at $q=0$, we need $\delta(q)$ to include the zero Bogoiulbov excitation.

For calculating the scattering rate of the dipolar BECs, we substitute the Bogoliubov transformation in Eq.~\ref{eq:Bog_transform} to the interaction Hamiltonian Eq.~\ref{eq:H_int2},
\begin{equation}
\begin{split}
\hat{H}_{int}&=g(k+q_i-q_f)\hat{c}_{k_f}(u_{q_i}\hat{b}_{q_i}+v_{-q_i}\hat{b}^\dag_{-q_i}+\delta(q_i))\\
&\times(u_{q_f}\hat{b}^\dag_{q_f}+v_{-q_f}\hat{b}_{-q_f}+\delta(q_f))
\end{split}
\end{equation}
and solve Fermi's Golden rule in Eq.~\ref{eq:Fermi Golden}. As described in the main text, the initial state of the transition in Eq.~\ref{eq:Fermi Golden} is the ground state in the quasi-particle field (\ket{q=0}) and photon field (\ket{k=0}). The transition creates one Bogoliubov excitation (\ket{q_f}) and one photon field excitation (\ket{k_f}). Eq. ~\ref{eq:Fermi Golden} becomes
\begin{equation}
    \Gamma_{i{\rightarrow}f}=\frac{2\pi}{\hbar}|\braket{q_f, k_f|\hat{H}_{int}}{q_0, k_0}|^2\delta(E_f-E_i).
\end{equation}
\paragraph*{\bf Calibration of Rayleigh scattering rate}
In our experiment, the Rayleigh scattering rate has been independently calibrated at different magnetic field.  Considering Lande g-factors for $\ket{J'=7;m'=-7}$ and $\ket{J=6;m=-6}$ are $g_e=-11.71$~MHz/G and $g_g=-9.77$~MHz/G, respectively, we compensate the change of detuning  $\delta$  induced by the Zeeman shift by adjusting the intensity $I$ of the pump beam following the relation $$R_{scatt}=\frac{\Gamma_{583}}{2}\frac{\Omega_{R}^2/2}{\delta^2+\Omega_{R}^2/2+\Gamma_{583}^2/4}$$ with the Rabi frequency $\Omega_{R}=\Gamma_{583}\sqrt{\frac{I}{2I_{sat}}}$. As an example, we select 300~mG and 400~mG in Fig.~\ref{figS1} where the phase fluctuation can be avoided with the small scattering length, and monitor the number of rocoiling atoms as a function of Rayleigh scattering rate. 

\begin{figure}
	\includegraphics[width=\linewidth]{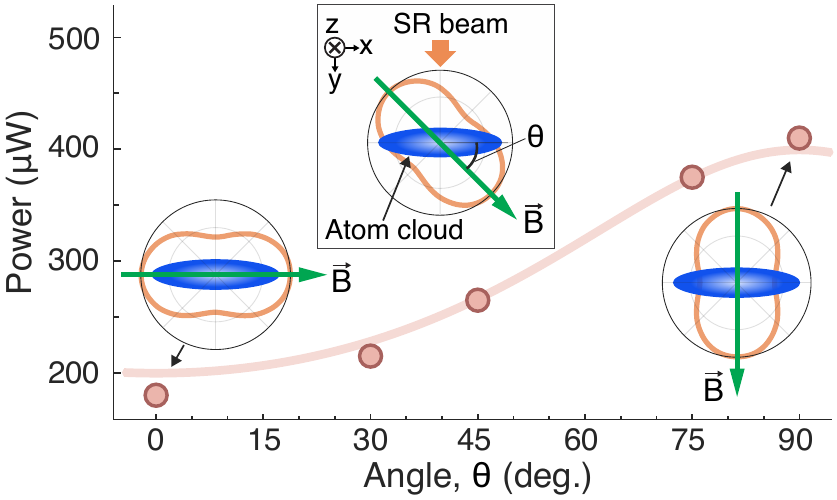}
	\caption{ {\bf Calibration of effective pumping intensity} When collective lights scattering occurs, we keep the emission rate along the x-direction constant as the magnetic field rotates. The required pump beam intensity scales as  $1/(1+\cos^2{\theta})$ as described by a solid curve.  Ellipses and dumbbells in insets are atom condensates and radiation patterns, respectively at 0$^{\circ}$, 45$^{\circ}$, and 90$^{\circ}$. Here, $\theta$ is an angle of magnetic field (green arrow) with respect to the longitudinal direction of the atom cloud. }
	\label{figS2}
\end{figure}

\vspace{20pt}
\paragraph*{\bf Calibration of effective pumping intensity}
Radiation from an atom has an angular dependence on a quantization axis. Therefore, the Rayleigh scattering rate changes while we rotate the external magnetic field during experiments. Additionally, only radiation along the longitudinal direction of the atom cloud is amplified by the gain medium and contributes the superradiance. Therefore, we calibrate the intensity of the pump beam such that spontaneous emission along the longitudinal direction of the atom cloud remains constant for the different quantization directions. Since the superradiance process involves $\sigma$-transitions, the angular distribution of emission spectrum $I_{eff}$ is given as $I_{eff}(\alpha)=I(1+\cos^2{\alpha})/2$ where $I$ is the intensity of the  pump beam and $\alpha$ is an angle of the emission direction with respect to the quantization axis. 

To observe asymmetric superradiance induced by the dipole-dipole interaction, we keep the constant scattering rate into the end- fire mode (i.e. $I_{eff}$) by using variable pump intensity $I(\theta)$ as $I(\theta)=2I_{eff}/(1+\cos^2{\theta})$ where the magnetic field has an angle $\theta$ with respect to the x axis. Fig.~\ref{figS2} shows the angular dependence of the required pump beam intensity at different angle of magnetic field.

\vspace{20pt}

\paragraph*{\bf Rotation of magnetic field}
Bias magnetic field was generated by three pairs of Helmholtz coils perpendicular to each other in Fig.~1. To rotate the direction of the magnetic field, we tuned currents in coils within 500~us. Due to the dipole-dipole interaction, the shape of dipolar condensates could be changed when the direction of the magnetic field was tuned. Moreover, the cloud shape change affects the gain of the superradiance. Hence, we prevented the cloud shape change by sending a pump beam immediately after rapidly rotating the magnetic field in less than 0.5~ms.

\begin{figure}
	\includegraphics[width=0.8\linewidth]{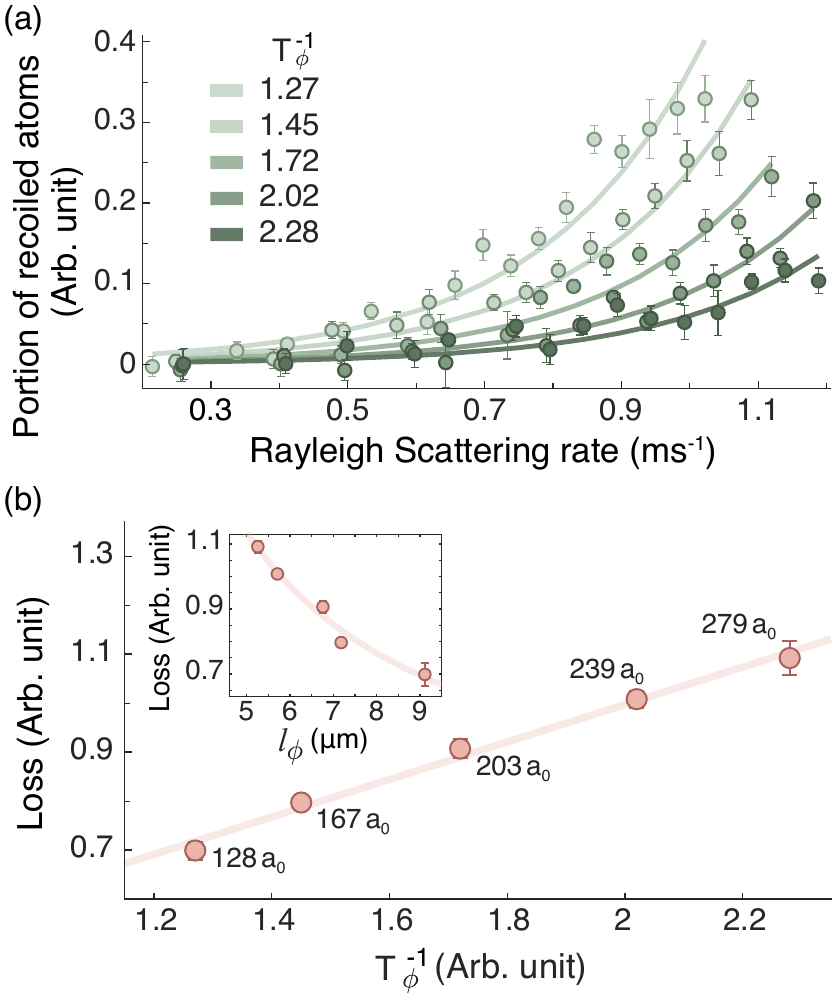}
	\caption{{\bf Control of superradiance threshold with the phase fluctuation} (a) Observation of portion of recoiled atoms versus the Rayleigh scattering rate $R_{sc}$ for different phase fluctuation being proportional to $T_\phi^{-1}$. To extract the threshold of Rayleigh scattering rate, corresponding to the loss term $L$, the data curve is fitted to the equation 
	$N=Ae^{(R_{sc}-L)t_{pulse}}$ where $t_{pulse}$=100$\mu$s and $A$ is arbitrary coefficient. Error bars indicate the standard error of 20 consecutive measurements. (b) $L$ is plotted against the $T_\phi^{-1}$ and $a_s$. The threshold of superradiance linearly increases with the amount of phase fluctuation in the condensate.   Inset illustrated the decrease of the $L$ against the increase of the coherence length $l_\phi$. Error bars indicate 90$\%$ confidence interval of loss for the fitting in (a).  Dashed lines are guides for the eyes. }
	\label{fig7}
\end{figure}

\vspace{20pt}

\paragraph*{\bf Control of superradiance threshold with the phase fluctuation}

When pump photons are scattered from a condensate, long-lived quasi-particles are excited leading to directional amplified Rayleigh scattering. The onset of the superradiant process is accounted for with the number of recoiling atoms $N$, following the rate equation $\dot{N}=(G-L)N$ where $G$ is the gain of the superradiance and $L$ is the loss term. A dipolar BEC of $^{168}$Er offers a new opportunity for the unprecedented control of superradiance. First, with increasing s-wave scattering length, phase fluctuation shortens the coherence length and enhances the decay of the matter-wave grating resulting in the increase of $L$. Secondly, in contrast to alkali atoms with contact interactions, a dipolar BEC  exhibits an anisotropic excitation spectrum due to the dipole-dipole interaction~\cite{Lahaye.2009,Bismut.2012,Wenzel.2018} resulting in anisotropic superradiant gain $G$ being sensitive to the external magnetic field direction. The condesnate sample, however, is macroscopic in our experiment, and thus the evolution of the matter wave grating in the presence of dipolar interaction may affect the loss term $L$. This effect could be further considered by examining semi-classical dynamics. 

In the experiments, we test for the control of the loss term $L$ by adjusting the phase coherence length of the condensate $
 l_{\phi}$. When the condensate begins to interact with the superradiance pump photons, recoiling atoms form a matter-wave grating with the lifetime $\tau_c$ being proportional to $\tau_c\propto\frac{l_{\phi}}{v}$ where $v$ and $l_{\phi}$ are the velocity of the recoiled atoms and the characteristic coherence length of the condensate, respectively. Here, the loss term increases with $l_{\phi}^{-1}$  as $L\propto 1/\tau_c\propto \frac{v}{l_{\phi}}$. For a condensate without phase fluctuations, the coherence length can be considered as the Thomas-Fermi radius $l$ of the condensate. With phase fluctuations, however, the coherence length is given by
$ l_\phi\simeq l(T_\phi / T)$
where $T$ is the temperature and $T_\phi$ is the characteristic temperature of phase fluctuation defined as $T_\phi=15(\hslash\omega_z)^2N_{tot}/32\mu\propto a_s^{-2/5}$ with the reduced Planck constant $\hbar$. Here,  $\mu$ is the chemical potential and the $a_s$ is the scattering interaction length. Therefore, the loss term can be controlled by the scattering length as $ L \propto T_{\phi}^{-1} \propto a_s^{2/5}$. In the experiment, we exploit the Feshbach resonance at $B_{res}\simeq$~912~mG and control $a_s$ from 128$a_0$ to 279$a_0$ with $T_{\phi}$ ranging from 100~nK to 50~nK where $a_0$ is the Bohr radius.

To control the scattering length $a_s$, the magnetic field is changed from 400~mG to the target value within $t_{ramp}$=2.7~ms right before the pump beam is switched on, while the field direction is kept along the x-axis (i.e. $\theta$=0). For different magnetic fields, we tune the Rayleigh scattering rate up to 1.2 ms$^{-1}$ based on the independent calibration, which effectively controls the gain term $G$. Subsequently, we record the portion of recoiled atoms against the change of Rayleigh scattering rate while we adjust the characteristic temperature of phase fluctuation ($T_{\phi}^{-1}$) as shown in Fig.~\ref{fig7}(a). Total recoiled atom number $N$ exponentially grows as the Rayleigh scattering rate (or $G$) increases. Then, we extract the loss term $L$ by fitting the initial rise of the recoiling atom number with exponential growth. We deduce the loss term from the measured threshold value. Fig.~\ref{fig7}(b) shows the change in the loss ($L$) for variable phase fluctuation characterized by $T_{\Phi}^{-1}$. The inset shows the decay of loss against the increase of coherence length. It is worth noting that the loss term $L$ may also be affected by other dephasing mechanisms, such as spontaneous Rayleigh scattering, incoherent collisions, or light-assisted collisions. However, within our experimental parameter range, the electronic structure remains almost unchanged, making such dephasing mechanisms negligible compared to the effect of phase fluctuation.


\begin{thebibliography}{0}%
\makeatletter
\providecommand \@ifxundefined [1]{%
 \@ifx{#1\undefined}
}%
\providecommand \@ifnum [1]{%
 \ifnum #1\expandafter \@firstoftwo
 \else \expandafter \@secondoftwo
 \fi
}%
\providecommand \@ifx [1]{%
 \ifx #1\expandafter \@firstoftwo
 \else \expandafter \@secondoftwo
 \fi
}%
\providecommand \natexlab [1]{#1}%
\providecommand \enquote  [1]{``#1''}%
\providecommand \bibnamefont  [1]{#1}%
\providecommand \bibfnamefont [1]{#1}%
\providecommand \citenamefont [1]{#1}%
\providecommand \href@noop [0]{\@secondoftwo}%
\providecommand \href [0]{\begingroup \@sanitize@url \@href}%
\providecommand \@href[1]{\@@startlink{#1}\@@href}%
\providecommand \@@href[1]{\endgroup#1\@@endlink}%
\providecommand \@sanitize@url [0]{\catcode `\\12\catcode `\$12\catcode
  `\&12\catcode `\#12\catcode `\^12\catcode `\_12\catcode `\%12\relax}%
\providecommand \@@startlink[1]{}%
\providecommand \@@endlink[0]{}%
\providecommand \url  [0]{\begingroup\@sanitize@url \@url }%
\providecommand \@url [1]{\endgroup\@href {#1}{\urlprefix }}%
\providecommand \urlprefix  [0]{URL }%
\providecommand \Eprint [0]{\href }%
\providecommand \doibase [0]{http://dx.doi.org/}%
\providecommand \selectlanguage [0]{\@gobble}%
\providecommand \bibinfo  [0]{\@secondoftwo}%
\providecommand \bibfield  [0]{\@secondoftwo}%
\providecommand \translation [1]{[#1]}%
\providecommand \BibitemOpen [0]{}%
\providecommand \bibitemStop [0]{}%
\providecommand \bibitemNoStop [0]{.\EOS\space}%
\providecommand \EOS [0]{\spacefactor3000\relax}%
\providecommand \BibitemShut  [1]{\csname bibitem#1\endcsname}%
\let\auto@bib@innerbib\@empty
\end{thebibliography}%


\begin{thebibliography}{34}

\bibitem{Dicke.1954je7}   R.~H. Dicke, {Physical Review} \textbf{93}, 99 (1954).

\bibitem{Gross.1982l7}   M. Gross and S. Haroche, {Physics Reports} \textbf{93}, 301 (1982).

\bibitem{Yoshikawa.2005}   Y. Yoshikawa, Y. Torii and T. Kuga, {Physical Review Letters} \textbf{94}, 083602 (2005).

\bibitem{Inouye.1999}   S. Inouye, A. Chikkatur, D.~M. Stamper-Kurn, J. Stenger, D. Pritchard and W. Ketterle, {Science} \textbf{285}, 571  (1999).

\bibitem{Stenger.1999rlo}   J. Stenger, S. Inouye, D. Stamper-Kurn, A. Chikkatur, D. Pritchard and W. Ketterle, { Applied Physics B} \textbf{69}, 347 (1999).

\bibitem{Inouye.19990n6}   S. Inouye, T. Pfau, S. Gupta, A.~P. Chikkatur, A. Gorlitz, D.~E. Pritchard and W. Ketterle, { Nature} \textbf{402}, 641 (1999).

\bibitem{Fallani.2005}   L. Fallani, C. Fort, N. Piovella, M. Cola, F.~S. Cataliotti, M. Inguscio and R. Bonifacio, { Physical Review A} \textbf{71}, 033612 (2005).

\bibitem{Deng.2010gfl}   L. Deng, E.~W. Hagley, Q. Cao, X. Wang, X. Luo, R. Wang, M.~G. Payne, F. Yang, X. Zhou, X. Chen et~al., { Physical Review Letters} \textbf{105}, 220404 (2010).

\bibitem{Lu.2011je5}   B. Lu, X. Zhou, T. Vogt, Z. Fang and X. Chen, { Physical Review A} \textbf{83}, 033620 (2011).

\bibitem{Kampel.2011}   N.~S. Kampel, A. Griesmaier, M.~P.~H. Steenstrup, F. Kaminski, E.~S. Polzik and J.~H. Muller, { Physical review letters} \textbf{108}, 090401 (2011).

\bibitem{Lopes.2014}   R. Lopes, A. Imanaliev, M. Bonneau, J. Ruaudel, M. Cheneau, D. Boiron and C.~I. Westbrook, { Physical Review A} \textbf{90}, 013615 (2014).

\bibitem{Dimitrova.2017}   I. Dimitrova, W. Lunden, J. Amato-Grill, N. Jepsen, Y. Yu, M. Messer, T. Rigaldo, G. Puentes, D. Weld and W. Ketterle, { Physical Review A} \textbf{96}, 051603 (2017).
  

\bibitem{Wang.2010f8xf}   P. Wang, L. Deng, E.~W. Hagley, Z. Fu, S. Chai and J. Zhang, { Physical Review Letters} \textbf{106}, 210401 (2010).

\bibitem{Slama.2007}   S. Slama, S. Bux, G. Krenz, C. Zimmermann and P.~W. Courteille, { Physical Review Letters} \textbf{98}, 053603 (2007).

\bibitem{Baumann.2010}   K. Baumann, C. Guerlin, F. Brennecke and T. Esslinger, { Nature} \textbf{464}, 1301 (2010).

\bibitem{Kessler.2014}   H. Kessler, J. Klinder, M. Wolke and A. Hemmerich, { Physical Review Letters} \textbf{113}, 070404 (2014).

\bibitem{Deng.2010} L. Deng, M. G. Payne, and E. W. Hagley, { Physical Review Letters} 104, 050402 (2010).

\bibitem{Clark.2017}   L.~W. Clark, A. Gaj, L. Feng and C. Chin, { Nature} \textbf{551}, 356 (2017).

\bibitem{Fu.2018i3f}   H. Fu, L. Feng, B.~M. Anderson, L.~W. Clark, J. Hu, J.~W. Andrade, C. Chin and K. Levin, { Physical Review Letters} \textbf{121}, 243001 (2018).

\bibitem{Wu.2019p5j}   Z. Wu and H. Zhai, { Physical Review A} \textbf{99}, 063624 (2019).

\bibitem{Kim.2021}   K. Kim, J. Hur, S. Huh, S. Choi and J.-y. Choi, { Physical Review Letters} \textbf{127}, 043401 (2021).

\bibitem{Muller.2016} J. H. Muller, D. Witthaut, R. l. Targat, J. J. Arlt, E. S. Polzik, and A. J. Hilliard, Journal of Modern Optics \textbf{63}, 1886 (2016).

\bibitem{Lahaye.2009}   T. Lahaye, C. Menotti, L. Santos, M. Lewenstein and T. Pfau, { Reports on Progress in Physics} \textbf{72}, 126401 (2009).

\bibitem{Bismut.2012}   G. Bismut, B. Laburthe-Tolra, E. Marechal, P. Pedri, O. Gorceix and L. Vernac, { Physical Review Letters} \textbf{109}, 155302 (2012).

\bibitem{Wenzel.2018}   M. Wenzel, F. Bottcher, J.-N. Schmidt, M. Eisenmann, T. Langen, T. Pfau and I. Ferrier-Barbut, { Physical Review Letters} \textbf{121}, 030401 (2018).


\bibitem{Gross.1982} M. Gross and S. Haroche, Phys. Rep. 93, \textbf{301} (1982).


\bibitem{Deng.2013} L. Deng, E.W. Hagley, R.Q. Wang and C.W. Clark, Optics and Photonics News, May issue, (2013) 

\bibitem{Chomaz.2022}   L. Chomaz, I. Ferrier-Barbut, F. Ferlaino, B. Laburthe-Tolra, B.~L. Lev and T. Pfau, { arXiv:2201.02672} ,  (2022).


\bibitem{Seo.2020}   B. Seo, P. Chen, Z. Chen, W. Yuan, M. Huang, S. Du and G.-B. Jo, { Physical Review A} \textbf{102}, 013319 (2020).

\bibitem{qrl}   B. Seo, Z. Chen, M. Huang, M.~K. Parit, Y. He, P. Chen and G.-B. Jo, { Journal of Korean Physical Society} \textbf{82}, 901 (2023).

\bibitem{He.2024}  Y. He, Z. Chen, H. Zhen, M. Huang, M. K. Parit, and G.-B. Jo  {arXiv:2403.18683} (2024).


\bibitem{qutip} J. R. Johansson, P. D. Nation, and F. Nori, Comp. Phys. Comm. 184, 1234 (2013)

\bibitem{Nozieres.1990}   P. Nozieres, The Theory of Quantum Liquids (1990).


\bibitem{petrov.2015} D.S. Petrov, { Physical Review Letters} \textbf{115}, 155302 (2015)

\bibitem{Ferrier.2016} Igor Ferrier-Barbut, Holger Kadau, Matthias Schmitt, Matthias Wenzel, and Tilman Pfau
Physical Review Letters 116, 215301 (2016).

\bibitem{chomaz.2016} L. Chomaz et al., Phys. Rev. X 6, 041039 (2016).

\bibitem{Huang.2023} M. Huang et al., In Preparation (2024)

\bibitem{Cardenas-Lopez.2023} S. Cardenas-Lopez, S. J. Masson, Z. Zager, and A. Asenjo-Garcia, { Physical Review Letters} 131, 033605 (2023).

\bibitem{Ticknor.2011}   C. Ticknor, R.~M. Wilson and J.~L. Bohn, { Physical Review Letters} \textbf{106}, 065301 (2011).

\bibitem{Lu.2012}   M. Lu, N.~Q. Burdick and B.~L. Lev, { Physical Review Letters} \textbf{108}, 215301 (2012).

\bibitem{Aikawa.2014}   K. Aikawa, A. Frisch, M. Mark, S. Baier, R. Grimm and F. Ferlaino, { Physical Review Letters} \textbf{112}, 010404 (2014).

\bibitem{Ni.2008}   K.~K. Ni, S. Ospelkaus, M.~H. G.~d. Miranda, A. Pe'er, B. Neyenhuis, J.~J. Zirbel, S. Kotochigova, P.~S. Julienne, D.~S. Jin and J. Ye, { Science} \textbf{322}, 231  (2008).



\end{thebibliography}
\end{document}